\documentclass[aps,pre,superscriptaddress,twocolumn,showpacs]{revtex4}
\usepackage{epsfig}

\newcommand{\be}{\begin{equation}}
\newcommand{\ee}{\end{equation}}
\newcommand{\bc}{\begin{center}}
\newcommand{\ec}{\end{center}}
\newcommand{\bi}{\begin{itemize}}
\newcommand{\ei}{\end{itemize}}
\newcommand{\ba}{\begin{eqnarray}}
\newcommand{\ea}{\end{eqnarray}}

\newcommand{\ignore}[1]{}
\newcommand{\mean}[1]{\left\langle #1 \right\rangle}

\begin{document}

\title{Conservation laws for the voter model in complex networks}

\author{Krzysztof Suchecki}
\affiliation{Instituto Mediterr\'aneo de Estudios Avanzados IMEDEA
(CSIC-UIB), E-07122 Palma de Mallorca, Spain} \affiliation{Center
of Excellence for Complex Systems Research and Faculty of Physics,
Warsaw University of Technology, PL-00-662 Warsaw, Poland}

\author{V\'{\i}ctor M. Egu\'{\i}luz}
\affiliation{Instituto Mediterr\'aneo de Estudios Avanzados IMEDEA
(CSIC-UIB), E-07122 Palma de Mallorca, Spain}
\email{victor@imedea.uib.es}

\author{Maxi San Miguel}
\affiliation{Instituto Mediterr\'aneo de Estudios Avanzados IMEDEA
(CSIC-UIB), E-07122 Palma de Mallorca, Spain}
\email{maxi@imedea.uib.es}

\date{\today}

\begin{abstract}
We consider the voter model dynamics in random networks with an
arbitrary distribution of the degree of the nodes. We find that
for the usual node-update dynamics the average magnetization is
not conserved, while an average magnetization weighted by the
degree of the node is conserved. However, for a link-update
dynamics the average magnetization is still conserved. For the
particular case of a Barabasi-Albert scale-free network the voter
model dynamics leads to a partially ordered metastable state with
a finite size survival time. This characteristic time scales
linearly with system size only when the updating rule respects the
conservation law of the average magnetization. This scaling
identifies a universal or generic property of the voter model
dynamics associated with the conservation law of the
magnetization.
\end{abstract}


\maketitle

\section{Introduction}

Conservation laws play an important role in the characterization
and classification of different nonequilibrium processes of
ordering dynamics. For example, in Kinetic Ising models one
distinguishes between Glauber and Kawasaki dynamics. In Glauber
dynamics the individual dynamical step is that of flipping a spin,
while in Kawasaki dynamics two nearest neighbor spins are
exchanged. Kawasaki dynamics conserves magnetization and Glauber
dynamics does not. As a consequence, the Glauber and Kawasaki
dynamics give rise to different scaling laws for domain growth in
coarsening processes \cite{Gunton83}, and they define different
nonequilibrium universality classes. The conservation law of
Kawasaki dynamics is implemented at each time step of a stochastic
dynamics. A different type of conservation law is the one that
refers to an ensemble average. An example of such conservation
laws is the conservation of the average spin (global
magnetization) in the voter model \cite{Liggett85,Marro}. When
studying spin dynamical models in regular lattices the existence
of an ensemble conservation law does not imply an elementary step
conservation such as imposed in the Kawasaki dynamics. Recent
interest in ordering processes focuses in situations in which the
spins are located in the nodes of a complex network, {\em i.e.} a
network with a large heterogeneity in the number of nearest
neighbors with which each spin interacts. This does not affect the
fulfillment of a conservation law of the type of the Kawasaki
dynamics, but the implementation of an ensemble average
conservation law requires a careful thought of the dynamical
rules. As an interesting example, we discuss in this paper the
differences between {\em node} and {\em link-update} dynamics for
a voter model in which spins are located in the nodes of a random
network with an arbitrary degree distribution. Only link-update
dynamics respects the global magnetization conservation law, while
another conservation law exists for node-update dynamics.

The standard voter model is defined by a set of ``voters" with two
opinions (spins $\sigma_i= \pm 1$) located in the nodes of an
hypercubic lattice. The elementary dynamical step consists in
randomly choosing one node (asynchronous update) and assigning to
it the opinion of one of its nearest neighbors, also chosen at
random (node-update). One time step corresponds to updating a
number of nodes equal to the system size, so that each node is on
average updated once. In $d=1$ this dynamics is equivalent to the
zero temperature Glauber kinetic Ising model. In general
dimensionality, the global magnetization is conserved in the
thermodynamic limit of large systems and the dynamics is dominated
by interfacial noise. The infinite system coarsens for $d \leq 2$,
with a slow logarithmic decay in the critical dimension $d=2$. At
variance with other ordering dynamics, coarsening takes place here
without surface tension. The role of the conservation law of the
magnetization and of the $Z_2$ symmetry ($\pm 1$ states) in the
voter dynamics universality class has been studied in detail in
the critical dimension $d=2$ of regular lattices \cite{Dornic01}.
We are here interested in situations in which there is no
long-time coarsening: for $d>2$ a finite system in a regular
lattice reaches one of the homogeneous attractors in which all the
$N$ spins have the same value. The time to reach such consensus
$\tau$, or survival time, scales as $\tau \propto N$, so that
there is no complete ordering in the thermodynamic limit
\cite{regularvoter}. This same scaling behavior has been also
found for the voter model in a small-world network
\cite{Castellano03,Vilone04} so that it can be identified as a
generic property of the voter model dynamics. We show in this
paper that for a Barabasi-Albert (BA) scale-free network
\cite{Barabasi} the scaling law $\tau \propto N$ is only obtained
when the average magnetization is conserved, that is when
link-update dynamics is used.

\section{Node vs link-update in the voter model}

Generically, in a complex network such as the small-world network,
there is an heterogeneous degree distribution with nodes having a
different number of links. In this case, as explained in more
detail below, node-update dynamics does not guarantee the
conservation of the average magnetization. The conservation is
guaranteed in a link-update dynamics in which the elementary
dynamical step consists in randomly choosing a pair of nearest
neighbor spins, {\it i.e.} a link, and randomly assigning to both
nearest neighbor spins the same opinion when they had different
opinion, and leaving them unchanged otherwise. The reported
simulations of the voter model in a small world network seem to
use a node-update \cite{Castellano03}, while the analytical
results are obtained in an approximation which enforces
conservation of the average magnetization \cite{Vilone04}. Since
in both cases the scaling law $\tau \propto N$ is obtained, the
role of the conservation law in this generic property is unclear.
Still, the degree heterogeneity in a small world network is rather
small, and therefore its effect on the conservation law for
node-update is probably not significant. The question is much more
crucial when considering, for example, a scale-free network which
has nodes with large degree heterogeneity, as we do next.

\section{Conservation laws}

\begin{figure}
\centerline{\epsfig{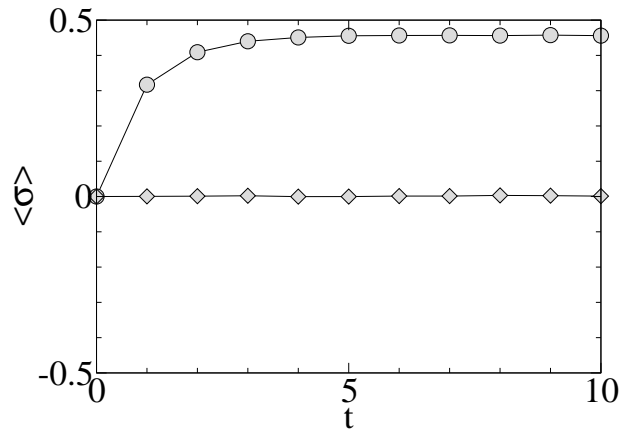}}
\caption{\label{ba_same} Ensemble average magnetization
$\mean{\sigma}$ in a BA network. The most connected half of the
nodes in the network have initial spin $\sigma_i=+1$, while the
other half has initial spin $\sigma_i=-1$. System size $N=1 000$,
average degree $\overline{k}=8$, and averaged over $1 000$
realizations. Circles: node-update; Diamonds: Link-update.}
\end{figure}

Results for the ensemble average normalized magnetization, \be
\mean{\sigma(t)} = \frac{\mean{\sum_{i=1}^N \sigma_i(t)}}{N}~,
\ee
of a voter model dynamics in a BA network \cite{Barabasi} are
shown in Fig.~\ref{ba_same} for node-update and link-update
dynamics. The ensemble average indicated as $\mean{\cdot}$ is an
average over the realizations of the stochastic dynamics and
different initial conditions. For both updating rules we have
considered the same initial distribution of spins in the initial
configuration, with half of the nodes with spin $+1$, and half
with spin $-1$. To provide more clear evidence of the dynamical
differences of the two update rules we have chosen an initial
configuration in which all the nodes that were given initial spin
$+1$ have a higher degree than those that were given initial spin
$-1$. For the node-update dynamics we find that the average spin
is not conserved, but rapidly changes towards the positive side,
due to the influence of the high-degree nodes in their immediate
neighborhood. For the link-update dynamics the average
magnetization remains at its vanishing initial value.

For any complex network with an heterogeneous distribution of the
degree of the nodes, the differences between the two updating
rules are easily understood as follows. The conservation of the
average magnetization in a regular lattice relies on the fact
that, if node $i$ and $j$ have different opinions and are
connected, the probability for $j$ to change to the spin of $i$ in
a time step of the dynamics is the same as for $i$ to change to
the spin of $j$. Since this is true for all nodes, the ensemble
average magnetization is conserved. However, in a network where
the nodes have not the same degree, this is not longer true for
node-update dynamics. For instance, if $i$ is a highly connected
node with degree $k_i$, and $j$ is a node with a low degree $k_j
<k_i$, and they have different spins, then the probability
$P_{ij}$ that $i$ changes to the spin of $j$ ($(Nk_i)^{-1}$) is
smaller than the probability $P_{ji}$ that $j$ changes to the spin
of $i$ ($(Nk_j)^{-1}$). Thus, the average magnetization is not
conserved. Choosing the node to be updated preferentially, so that
$P_{ji} = P_{ij}$, makes the average spin conserved again.
Preferentially choosing the node to be update in this way is
equivalent to randomly choosing a link in the network and updating
it in random direction (link-update).

As we have argued the ensemble average normalized magnetization
$\mean{\sigma}$ is not conserved in a complex network in the voter
model with node-update rule. In order to compensate for the
different degrees of the nodes, we consider a degree weighted
normalized magnetization \cite{Bianconi,Agata}
\be \Sigma (t)=
\frac{\sum_{i=1}^{N} k_i \sigma_i(t)} {\sum_{i=1}^{N} k_i}~, \ee
where $k_i$ is the degree of the node $i$. The total number of
links is introduced in the denominator to normalize the weighted
magnetization between $[-1,1]$.

If we define $S(t)= \sum_{i=1}^{N} k_i \sigma_i(k,t)$, then for a
given configuration and averaging over stochastic realizations,
the expected change $\mean{\Delta S_{ij}}_c$ in a time step of the
dynamics due to node $i$ changing spin to the spin of node $j$ is
given by \be \mean{\Delta S_{ij}}_c = \frac{(\sigma_j(t) -
\sigma_i(t))}{Nk_i}k_i~, \ee while \be \mean{\Delta S_{ji}}_c =
\frac{(\sigma_i(t) - \sigma_j(t))}{Nk_j}k_j~, \ee where
$\mean{\cdot}_c$ represents an average over stochastic
realizations for a given configuration $c$ at time $t$. Thus
$\mean{\Delta S_{ij}+\Delta S_{ji}}_c=0$. As this argument applies
for any pair of neighbors, and any configuration, {\em the
ensemble average weighted magnetization $\mean{\Sigma}$ is
conserved for the voter model using node-update in complex
networks with arbitrary degree distributions.}

\begin{figure}
\centerline{\epsfig{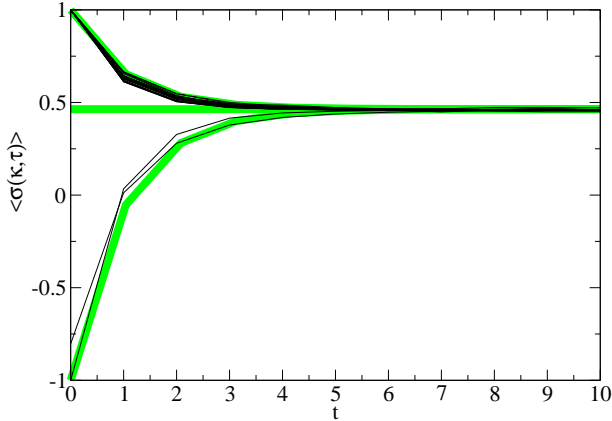}}
\caption{\label{sk_ba} Time evolution of $\mean{\sigma(k,t)}$ for
node-update dynamics in a BA network with $\overline{k}=8$.
Initial configuration as in Fig.~\ref{ba_same}. System size
$N=10,000$, average degree $\overline{k}=8$ and average taken over
$10,000$ realizations. Thin lines correspond to numerical data;
thick lines correspond to the analytical predictions
(Eq.~(\ref{sigmak})) for the initial values $\sigma(k,0)= -1$ and
$+1$.}
\end{figure}

Given the conservation law of $\mean{\Sigma}$ we can find
the asymptotic value of the average magnetization. To this end, we
introduce the normalized magnetization of the nodes with given degree $k$ as:
\be
\sigma(k,t)= \frac{\sum_{i:k_i=k} \sigma_i(t)}{N_k}~,
\ee
where $N_k$ is the number of nodes with degree $k$ and the sum in
the numerator is over all nodes with the same degree $k$.

For a given configuration $c$, the expected change of spin of a node
$i$ with degree $k_i$ due to interaction with its neighbors in a
time step of the dynamics is
given by
\be
\mean{\Delta \sigma_i (t)}_c = \sum_{j\in{\cal V}_i}
\frac{\sigma_j (t) -\sigma_i (t)}
{k_i}~,
\ee
where ${\cal V}_i$ is the neighborhood of node $i$, that is the
nodes connected by a link to node $i$. From this expression, if we
add for all the nodes with the same degree we obtain
\be
\mean{\Delta \sigma(k,t)} = \sum_{i:k_i=k}\sum_{j\in{\cal V}_i}
\frac{\sigma_j(t)-\sigma_i(t)}{N_kk_i}~.
\label{prev}
\ee
We can now split the r.h.s of Eq.~(\ref{prev}) in two terms. The second term is
simply $\sigma(k,t)$. For the first term, we assume a mean field
approximation, {\it i.e.}, we consider a random network where
the sum over neighbors is equivalent to a random sampling over the
whole network. Then
\be
\sum_{i:k_i=k}\sum_{j\in{\cal V}_i}\frac{\sigma_j(t)}{N_kk_i}
= \frac{\sum_{k} P(k)k \sigma(k,t)}{\sum_{k} P(k)k}
= \frac{\sum_{i=1,N} k_i \sigma_i(t)}{\sum_{i=1,N} k_i}
= \Sigma(t)~,
\ee
where $P(k)=N_k/N$ is the degree distribution of the network, that
is, the probability to find a node of degree $k$.

Thus, $\mean{\Delta \sigma(k,t)}_c = \Sigma - \sigma(k,t)$, where
$\Sigma(t)$ and $\sigma(k,t)$ are calculated in the given
configuration $c$. Averaging over different configurations, we
find the evolution equation of the ensemble average of
$\sigma(k,t)$ \cite{Huberman}:
\be \frac{d \mean{\sigma(k,t)}}{dt}
= \mean{\Sigma} - \mean{\sigma(k,t)}~.
\label{sigma} \ee

\begin{figure}
\centerline{\epsfig{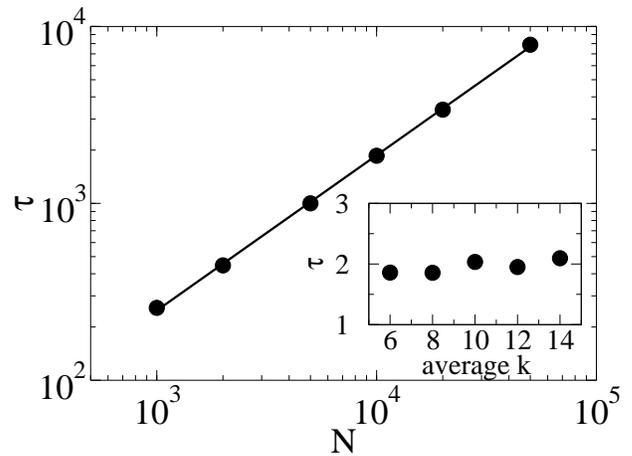}}
\caption{\label{ba_inset_T} Survival time for node-update dynamics
in BA networks of different sizes $N$, and average degree
$\overline{k}=6$. Inset: Survival times for different average
degree. All data obtained from at least $1000$ realizations of
networks of size $10000$.}
\end{figure}

Given the conservation law for $\mean{\Sigma}$, the solution of
Eq.~(\ref{sigma}) is an exponential approach to to the asymptotic
value
\be
\mean{\sigma(k,t)} = (\mean{\sigma(k,0)} - \mean{\Sigma}) e^{-t}
+  \mean{\Sigma}~.
\label{sigmak}
\ee
In the long time limit
$\mean{\sigma(k,t)}$ approaches a constant value $\mean{\Sigma}$
independent of $k$. Therefore this constant value coincides with
the long time limit of the ensemble average normalized
magnetization $\mean{\sigma (t \to \infty)}$. This value is a property
of the ensemble of initial configurations.

A numerical check of these general results for the particular case
of the BA network is shown in Fig.~\ref{sk_ba} where the fast
exponential decay to the final average value is shown. The
asymptotic value corresponds to the analytical prediction of
$\mean{\Sigma}$ calculated in the initial configurations, that is,
the ensemble of initial configurations.

\section{Survival times in Barab\'asi-Albert scale-free networks}

We next study the consequences of the two different updating rules
and associated conservation laws in the ordering dynamics of the
voter model in a BA network. For any of the two updating rules the
system falls, after an initial transient, in a metastable
partially ordered state until a finite system size fluctuation
takes the system to one of the two ordered attractors of the
dynamics. This qualitative behavior is similar to the one found in
a small world network \cite{Castellano03} and like wise can be
characterized in terms of the temporal evolution of the average
interface density $\rho$, defined as the density of links
connecting sites with different opinions. In a given realization
of the dynamics $\rho$ initially decreases indicating a partial
ordering of the system. After this initial transient $\rho$
fluctuates randomly around an average value until a fluctuation
orders the system leading to an absorbing state with $\rho=0$.
Considering an ensemble of realizations, the ordering of each of
them happens randomly with a constant rate. This is reflected in
an exponential decay of the average interface density
\be
\mean{\rho} \propto e^{-\frac{t}{\tau}}~,
\ee
where $\tau$ is the survival time of the partially ordered
metastable state. This survival time turns out to be a quantity
that diverges with growing system size $N$, so that the system
does not order in the thermodynamic limit.

\begin{figure}
\centerline{\epsfig{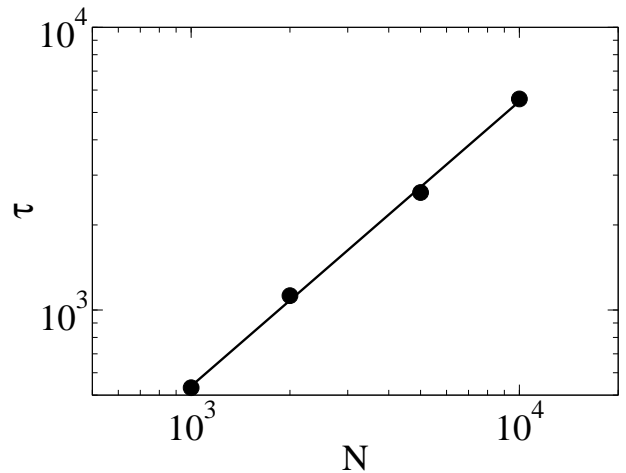}}
\caption{\label{ba_prefN} Survival time for link-update dynamics
in BA networks of different sizes $N$, and average degree
$\overline{k} =8$. All data obtained from $1000$ realizations.}
\end{figure}

We have measured the characteristic time $\tau$ for the two
updating dynamical rules and for different system sizes $N$ and
different mean degree $\overline{k}$. Our results are summarized
in Figs.~\ref{ba_inset_T} and \ref{ba_prefN}. For the node-update
rule (Fig.~\ref{ba_inset_T}), in which there is no conservation
law of the average magnetization, we have found that $\tau$ scales
with system size $N$ as
\be
\tau \propto N^\gamma~,
\ee
where $\gamma = 0.88 \pm 0.01$. For a fixed system size, the value
of the average degree $\overline{k}$ does not seem to have any
definite influence on $\tau$ (see the inset of
Fig.~\ref{ba_inset_T}). The value of the exponent $\gamma$ is
consistently and significantly different from $\gamma=1$ which is
the exponent analytically found for regular hypercubic lattices
and for an annealed small world network \cite{Vilone04}. In these
last two cases the dynamical rules respect the conservation of the
average magnetization. When we implement the link-update rule
which also conserves the average magnetization we find a result
for $\gamma$ which is consistent with $\gamma=1$
(Fig.~\ref{ba_prefN}).

\section{Conclusions}

In summary, we have shown that the voter model dynamics does not
lead to an ordered state in a scale-free network in the
thermodynamic limit. This is consistent with the results for a
small world network, and in general for networks of dimensionality
$d>2$. Finite size effects order the system in a time which
depends on the updating dynamical rule. Only for the updating rule
fulfilling a conservation of the global magnetization this time
scales linearly with the system size. This is also consistent with
the result for regular hypercubic lattices of $d>2$ and for the
voter model in annealed small-world networks \cite{Vilone04}. Such
scaling can then be taken as a proper characterization of
universal properties of the dynamics of the voter model.

We note that there are several instances in which the conservation
law of the global magnetization is naturally broken independently
of the updating rule, as for example the consideration of a zealot
\cite{zealot}, or the dynamics in a directed network. On the other
hand there are spreading phenomena with "spins" having $N>2$
states which show voter-like generic properties in d=2 regular
lattices and that fulfill the global magnetization conservation
law \cite{szabo}. Further study of these other cases and the
implications of other conservation laws, as the one reported here
for node-update dynamics, would be useful to identify generic and
nongeneric properties of the voter model dynamics.

\acknowledgments We acknowledge financial support from MCYT
(Spain) through project CONOCE (BFM2000-1108). KS thanks Prof.
Janusz Holyst for very helpful comments.



\begin{thebibliography}{0}

\bibitem{Gunton83}
  Gunton J.D., San Miguel M., Sahni P.S.
  Phase Transitions and Critical Phenomena
  Vol. 8, pp 269--466, C. Domb, J. Lebowitz Eds, Academic Press, London
  (1983).


\bibitem{Liggett85}
  Liggett T.M.
  Interacting Particle Systems, Springer, New York (1985).

\bibitem{Marro}
  Marro J., Dickman R.
  Nonequilirium Phase Transitions in Lattice Models,
  Cambridge University Press, Cambridge (1999).

\bibitem{Barabasi}
  Barab\'asi A.-L., Albert R. Science {\bf 286}, 509--512 (1999).

\bibitem{Dornic01}
  Dornic I., Chate H., Chave J.,
  Hinrichsen H. Phys. Rev. Lett. {\bf 87}, 045701 (2001).

\bibitem{regularvoter}
  Frachenbourg L., Krapivsky P.L. Phys. Rev. E {\bf53}, 3009 (1996).

\bibitem{Castellano03}
  Castellano C., Vilone D., Vespignani A.
   Europhysics Letters {\bf 63}, 153 (2003).

\bibitem{Vilone04}
  Vilone D., Castellano C.
   Phys. Rev. E {\bf 69}, 016109 (2004).

\bibitem{Bianconi}
  Bianconi C.
   Phys. Lett. A {\bf 313}, 166 (2002).

\bibitem{Agata}
  Aleksiejuk A., Holyst J.A., Stauffer D.
   Physica A {\bf 310}, 260 (2002).

\bibitem{Huberman}
  An independent related discussion of the dynamics of the average magnetization under node-update
  in a random network has been simultaneously reported by F. Wu and B. A. Huberman
  {\tt cond-mat/0407252}.

\bibitem{zealot}
  Mobilia M.
   Phys. Rev. Lett. {\bf 91}, 028701 (2003).

\bibitem{szabo}
  Ravasz M., Szabo G., Szolnoki A.
   Phys. Rev. E {\bf 70}, 012901 (2004).

\end{thebibliography}
\end{document}